\documentclass[onecolumn,aps,nofootinbib,preprint]{revtex4}
\pdfoutput=1

\usepackage{graphicx}
\usepackage{amsmath}
\usepackage{hyperref}
\usepackage{amssymb}
\usepackage{color}

\newcommand{\beq}{\begin{equation}}
\newcommand{\eeq}{\end{equation}}
\newcommand{\bea}{\begin{eqnarray}}
\newcommand{\eea}{\end{eqnarray}}

\newcommand{\ie}{\textit{i.e.}\ }

\newcommand{\refeq}[1]{Eq.~(\ref{#1})}

\newcommand{\ignore}[1]{}

\makeatletter
\def\simgt{\mathrel{\lower2.5pt\vbox{\lineskip=0.5pt\baselineskip=0pt
           \hbox{$>$}\hbox{$\sim$}}}}
\def\simlt{\mathrel{\lower2.5pt\vbox{\lineskip=0.5pt\baselineskip=0pt
           \hbox{$<$}\hbox{$\sim$}}}}
\makeatother

\begin{document}
\setlength{\unitlength}{1mm}

\title{Evidence for dark matter interactions in cosmological precision data?}

\author{Julien Lesgourgues$^a$}\email{Julien.Lesgourgues@physik.rwth-aachen.de}
\author{Gustavo Marques-Tavares$^{bc}$}\email{gusmt@stanford.edu}
\author{Martin Schmaltz$^b$}\email{schmaltz@bu.edu}
\affiliation{$^a$Institut f\"{u}r Theoretische Teilchenphysik und Kosmologie (TTK), RWTH Aachen University, 52056 Aachen, Germany}
\affiliation{$^b$Physics Department, Boston University, Boston, MA 02215, USA\\ }
\affiliation{$^c$Stanford Institute for Theoretical Physics, Department of Physics, Stanford University, Stanford, CA 94305\\ }

\begin{abstract}
\vskip.1in
\
We study a two-parameter extension of the cosmological standard model $\Lambda$CDM in which cold dark matter interacts with a new form of dark radiation. The two parameters correspond to the energy density in the dark radiation fluid $\Delta N_\mathrm{fluid}$ and the interaction strength between dark matter and dark radiation. The interactions give rise to a very weak ``dark matter drag" which damps the growth of matter density perturbations throughout radiation domination, allowing to reconcile the tension between predictions of large scale structure from the CMB and  direct measurements of $\sigma_8$. We perform a precision fit to Planck CMB data, BAO, large scale structure, and direct measurements of the expansion rate of the universe today. Our model lowers the $\chi$-squared relative to  $\Lambda$CDM by about 12, corresponding to a preference for non-zero dark matter drag by more than $3 \sigma$.
Particle physics models which naturally produce a dark matter drag of the required form include the recently proposed non-Abelian dark matter model in which the dark radiation corresponds to massless dark gluons.

\end{abstract}

\maketitle

\section{Introduction}

Cold dark matter (DM) has long been an important ingredient of the cosmological standard model $\Lambda$CDM. Evidence for its gravitational effects exists on a range of length scales from galaxy rotation curves and gravitational lensing to large scale structure (LSS) and the cosmic microwave background (CMB). Particle physics models of DM generically also predict non-gravitational interactions of DM. Such interactions could be responsible for determining the DM abundance via thermal freeze-out and would alter the clumping of DM at small scales. 
Until now all attempts to observe non-gravitational interactions of DM more directly have only yielded upper bounds. The searches include ``direct detection" of DM collisions with a target in the laboratory, ``indirect detection" of DM annihilation in regions of high DM density in the spectrum of photons emanating from such regions, and ``collider searches" which seek to produce and observe DM in particle collisions in the form of missing energy.

Another possible route to learning about the nature of DM is through precision cosmological measurements. Precision fits to the CMB, large scale structure, and several probes of the background expansion based on baryon acoustic oscillations (BAOs), supernovae luminosity, measuring the current expansion rate $H_0$, etc., are sensitive to detailed properties of the dark matter. For example, hot dark matter in the form of active neutrinos may not contribute to more than 1.9\% of the total dark matter density today, which turns into a bound on the total neutrino mass $\Sigma m_\nu<0.21$~eV (95\%CL)~\cite{Planck:2015xua} -- or $\Sigma m_\nu<0.12$~eV when using also Lyman-$\alpha$ forest data from quasar spectra~\cite{Palanque-Delabrouille:2015pga}. The dominant component is compatible with the assumption of cold dark matter (CDM), or possibly warm dark matter (WDM), but with strong bounds on the  velocity dispersion of warm particles in order to avoid a cut-off in the matter power spectrum on scales where precise data are available~\cite{Viel:2013fqw}. Similarly, consistency of the global fit limits the density of additional particles which may contribute to cosmic radiation to an equivalent number of neutrino species $N_\mathrm{eff}<3.7$ (95\%CL), while their masses satisfy model-dependent bounds, of order $m_x \leq 0.38\, (T_\nu/T_x)^3$~eV (95\%CL) for one thermal species $x$~\cite{Planck:2015xua}. Recent Planck data also requires standard neutrinos to be free-streaming, $c_{viscosity}^2=0.331\pm 0.037$ (68\%CL)~\cite{Planck:2015xua}. 

The ``cosmic concordance" $\Lambda$CDM model is impressively successful and some of its parameters have now been measured to permille accuracy. However, with the increase in precision of the measurements, there have also been indications in the data for effects which are not well described within $\Lambda$CDM. Perhaps most significantly, there is tension between the value of $\sigma_8$ predicted within $\Lambda$CDM with parameters fit from the CMB and BAO, and the value of $\sigma_8$ from more direct measurements of LSS with various techniques (CMB lensing, galaxy weak lensing, cluster mass function, etc.) This tension could be caused by systematics in the astrophysical data, or may point instead to new physics affecting dark matter, because $\sigma_8$ (roughly speaking the amplitude of matter fluctuations at scales of 8 Megaparsec) is predominantly determined by the growth of fluctuations in the DM component, during matter and radiation domination. By taking into account neutrino masses, or by introducing extra massive relics, it is possible to fit CMB and BAO data with smaller $\sigma_8$ values, but this is done at the expense of increasing the tension with measurements of $H_0$ from local redshift data, and of degrading the agreement between the matter density fraction $\Omega_m$ inferred from CMB data and from other cosmic probes~\cite{Planck:2015xua}.

In this paper we take the ``$\sigma_8$-problem" seriously and interpret the discrepancy as evidence for new physics affecting the DM. In particular, we investigate whether the tension between the CMB fit and the direct measurements of $\sigma_8$ may be resolved by including a very weak drag force between DM and radiation, which acts during radiation domination. Such an interaction would dampen the growth of density perturbations in the DM fluid, and therefore reduce the predicted matter power spectrum.

There have been several previous investigations of modifications to the matter power spectrum in the presence of new DM interactions, e.g. Refs.~\cite{Feng:2009mn,Serra:2009uu,Aviles:2011ak,Cyr-Racine:2013fsa,Dvorkin:2013cea,Wilkinson:2013kia,Wilkinson:2014ksa}. In these setups, the new interactions tend to be very strong at some scales and negligible at others, leading to threshold features in the power spectrum -- similar to WDM models. The data disfavor such features and put strong bounds on the interaction parameters, leaving little room for a reduction of the power spectrum at 8~Mpc scales. There have also been attempts to solve the problem with decaying dark matter (DDM)~\cite{Wang:2012eka,Enqvist:2015ara,Berezhiani:2015yta}. In that case, the strongest constraints come from CMB data: in order to avoid a huge ``late integrated Sachs-Wolfe'' effect, the DM lifetime must be so large (see e.g.~\cite{Audren:2014bca}) that $\sigma_8$ can only be slightly reduced. Hence, current DDM studies do not find very significant evidence for a non-zero decay rate. 

One way to solve the $\sigma_8$-problem without introducing other tensions is to consider a drag force which is both very weak and which results in a momentum transfer rate between the DM and radiation which scales with temperature in the same way as the Hubble expansion rate during radiation domination, i.e. proportional to $T^2$. A model which predicts this kind of drag force and damping of density perturbations was recently proposed in \cite{Buen-Abad:2015ova} (BMS). With such a scaling, the drag can act equally on density perturbations which enter the horizon at different times during radiation domination. This leads to a smooth suppression of the cold dark matter power spectrum at all scales during radiation domination. Since the growth of matter fluctuations is primarily modified during radiation domination (rather than matter domination), there is no enhancement of the late integrated Sachs-Wolfe effect in such models, and CMB data do not provide strong bounds on the interaction parameter. In Section~\ref{sec:equations}, we make this suggestion concrete and define the drag force, calculate the resulting momentum transfer rate, and discuss its qualitative effects on matter perturbations.

In order to add an interaction for the DM with radiation, it would appear most economical to couple DM to either photons or neutrinos. However, in both cases, it is difficult to get the desired scaling with temperature. For example, a momentum transfer rate proportional to $T^2$ is not compatible with Compton scattering off photons which scales as $T^4$, Coulomb scattering off electrons which scales as $T^{3/2}$ or weak interaction scattering off neutrinos which scales as $T^6$. In addition, any new interactions of especially photons but also neutrinos are tightly constrained by the global fit to the CMB~\cite{Serra:2009uu,Wilkinson:2014ksa,Wilkinson:2013kia}. Following the concrete model in BMS we therefore introduce a new self-interacting ``dark radiation" (DR) component to the energy density of the universe which may be described as a perfect fluid. The drag force arises from scattering between DR particles and the DM. Thus our proposal is to generalize $\Lambda$CDM with two parameters. One corresponds to the energy density in the fluid describing the new radiation component (parameterized and normalized like an effective number of extra neutrinos $\Delta N_\mathrm{fluid}$), and the other is the DM-DR interaction rate $\Gamma_0$ (this parameter gives the interaction rate today, but since its scaling with temperature is known, it characterizes the strength of the new drag force at any time).

The precise definitions of these parameters and their effect on the CMB and matter power spectrum are given in Section~\ref{sec:equations}. In Section~\ref{sec:models}, we review the example proposed by BMS of a particle physics model with a non-Abelian dark gauge group, and summarize the calculation of the momentum transfer rate. We also offer an alternative model with a massless dark photon coupled to the DM and massless fermions. Section~\ref{sec:fit} contains the main results of this paper. We perform a fit to CMB, LSS and BAO data and find a strong preference for non-zero interactions between the DM and DR. The best-fitting models have $\Gamma_0 \simeq 1.6 \times 10^{-7}~\mathrm{Mpc}^{-1} \simeq  1.6 \times 10^{-21}~\mathrm{s}^{-1}$ and $\Delta N_\mathrm{fluid}<0.7$, and the minimum $\chi^2$ improves by 11.4 over that of the minimal $\Lambda$CDM model with $\Delta N_\mathrm{fluid} =0$ and $\Gamma_0=0$ . The fit shows a $\sim3.7\sigma$ preference for a non-zero value of the drag coefficient. A non-vanishing DM drag of the kind that we are proposing is also found to be compatible with large values of the Hubble rate, as measured for instance by~\cite{Riess:2011yx,Freedman:2012ny} (when including such data, the minimum $\chi^2$ improves by 12.7).

\section{Generalizing $\Lambda$CDM with dark radiation and dark matter drag\label{sec:equations}}

\subsection{Modified cosmological perturbation equations}

We propose to add a new component of self-interacting dark radiation to $\Lambda$CDM. The radiation is comprised of relativistic particles with a self-scattering rate which is fast compared to the Hubble rate during radiation domination. Such radiation can be described as a perfect fluid with speed of sound $c_\mathrm{s}^2=1/3$ and no viscosity.

We assume that the dark radiation was in thermal equilibrium with the Standard Model particles early in the evolution of the universe, and that it decoupled during freeze-out of the dark matter. Then the temperature of the dark radiation today will be of the same order as the photon temperature. The photon temperature is expected to be higher because the photons inherit the entropy of heavier particles in the Standard Model during their freeze-out. The main effect of the DR (other than providing drag to the DM) is to contribute to the expansion rate of the universe, because it increases its average energy density. We choose to parameterize the energy density in the DR in analogy to an effective equivalent number of neutrino species
\bea
\Delta N_\mathrm{fluid}=N_\mathrm{dr}\, \left(\frac{T_\mathrm{dr}}{T_\nu}\right)^4 \times
\left\{ \begin{array}{ll} \frac87 & ~~{\rm (bosonic\ DR),} \\ 1 & ~~{\rm (fermionic\ DR).} \end{array} \right. 
\eea
Motivated by the specific particle physics models described in Section~\ref{sec:models} we will study values of $\Delta N_\mathrm{fluid}$ ranging from 0.07 to 1 (see section~\ref{subsec:data}).

The dark matter in our model is comprised of particles which become non-relativistic long before matter-radiation equality and therefore have negligible kinetic energy density. We take the dark matter to have negligible self-interactions and parameterize its contribution to the energy budget of the universe with $\omega_\mathrm{dm}=\Omega_\mathrm{dm} h^2$ as usual. 

The drag force between the DM and DR can be parametrized by the linear coefficient $\Gamma$ of friction which a non-relativistic DM particle of velocity $v$ experiences as it propagates through the thermal bath of radiation
\bea
\dot{\vec{v}} = -a \Gamma\, \vec{v} \ ,
\label{eq:tau-definition}
\eea
where the dot here and in the following represents a derivative with respect to conformal time and $a$ is the scale factor. The coefficient $\Gamma$ depends on the temperature of the DR. In any specific model it can be computed (see Section 3) from the rate of momentum transfer due to collisions of the DM particle as it travels though the DR. For the case of interest - CDM - the velocities in \refeq{eq:tau-definition} are non-relativistic and satisfy $c^2 \gg v^2 \gg T_\mathrm{dr}/M$ where $c$ is the speed of light and $M$ is the DM mass.  

We use the formalism of Ma and Bertschinger \cite{Ma:1995ey} and write the coupled evolution equations for density and velocity perturbations of the DM, $\delta_\mathrm{dm}$ and $\theta_\mathrm{dm}$, and the DR, $\delta_\mathrm{dr}$ and $\theta_\mathrm{dr}$. 
In Conformal Newtonian Gauge, the equations for the DM and DR overdensities in Fourier space are
\bea
\dot \delta_\mathrm{dm} & = & -\theta_\mathrm{dm} + 3 \dot \phi \\
\dot \theta_\mathrm{dm} & = &- \frac{\dot a}{a} \theta_\mathrm{dm} + a \Gamma (\theta_\mathrm{dr} - \theta_\mathrm{dm} ) + k^2 \psi \\
\dot \delta_\mathrm{dr} & = &- \frac{4}{3} \theta_\mathrm{dr} + 4 \dot \phi \\ 
\dot \theta_\mathrm{dr} & = & k^2 \frac{\delta_\mathrm{dr}}{4} + k^2 \psi + \frac{3}{4}\frac{\rho_\mathrm{dm}}{\rho_\mathrm{dr}} a \Gamma (\theta_\mathrm{dm} - \theta_\mathrm{dr}) \ .
\eea
Here $\rho_\mathrm{dm}$ and $\rho_\mathrm{dr}$ are the average energy densities of DM and DR, respectively; and $\phi$ and $\psi$ are the scalar metric perturbations in Conformal Newtonian gauge. Notice the absence of all higher moments of the DR perturbations, they vanish for perfect fluids.  

The purpose of displaying these equations is to draw attention to the drag terms proportional to $\Gamma$ which couple the two velocity equations. They represent the drag forces which result from collisions between the particles in the two fluids. To gain a rough understanding of what these terms do, note that during radiation domination the coefficient in the equation for the dark radiation is suppressed by the small ratio $\rho_\mathrm{dm}/\rho_\mathrm{dr}$. Therefore the main effect is the drag due to the DR on the DM. Note that the clock in these equations is set by the Hubble rate, which scales like $T^2$ during radiation domination. Therefore, if we require that the effect of the drag term is small at any instant, but uniform over a long interval of time, it must also scale as $T^2$. We will assume that this is the case, and show how to motivate this behavior with concrete particle physics models in Section~\ref{sec:models}.

After matter-radiation equality, the Hubble rate decreases more slowly, proportional to $T^{3/2}$, whereas the drag continues to be proportional to $T^2$. Thus the effects of the drag become less important after equality. We use $\Gamma_0$ to denote the value of the drag coefficient extrapolated to today. The drag coefficient at any other temperature is then
\bea
\Gamma = \Gamma_0 \left(\frac{T}{T_0}\right)^2 \ ,
\label{eq:tau-zero}
\eea
where $T_0=2.7255$ K is the current CMB temperature. 

\subsection{Effects on the CMB and LSS spectrum}

We implemented the above model in the Boltzmann code {\sc class}\footnote{{\tt github.com/lesgourg/class\_public} or {\tt class-code.net}}~\cite{Lesgourgues:2011re,Blas:2011rf}. Very few modifications of the public version of the code are required for this model. We implemented the new equations in both the Newtonian and Synchronous gauge, and checked that we get exactly the same results in the two gauges. The only difference is that in the Newtonian gauge, we can run with a density of ordinary non-interacting cold dark matter $\omega_\mathrm{cdm}$ set to exactly zero, while in the synchronous gauge we must set it to a negligible but non-zero value, e.g. $\omega_\mathrm{cdm}=10^{-10}$, since the latter gauge is by definition comoving with the CDM component. The code assumes natural units ($c=1$) and expresses conformal time and Fourier wavenumbers in Megaparsecs (Mpc). Hence $\Gamma_0$ is naturally expressed in inverse Mpc. It can be converted to inverse second by multiplying by $0.97 \times 10^{-14}\,$Mpc/s.

\begin{figure}[thb]
\centering
\includegraphics[width=.7\textwidth]{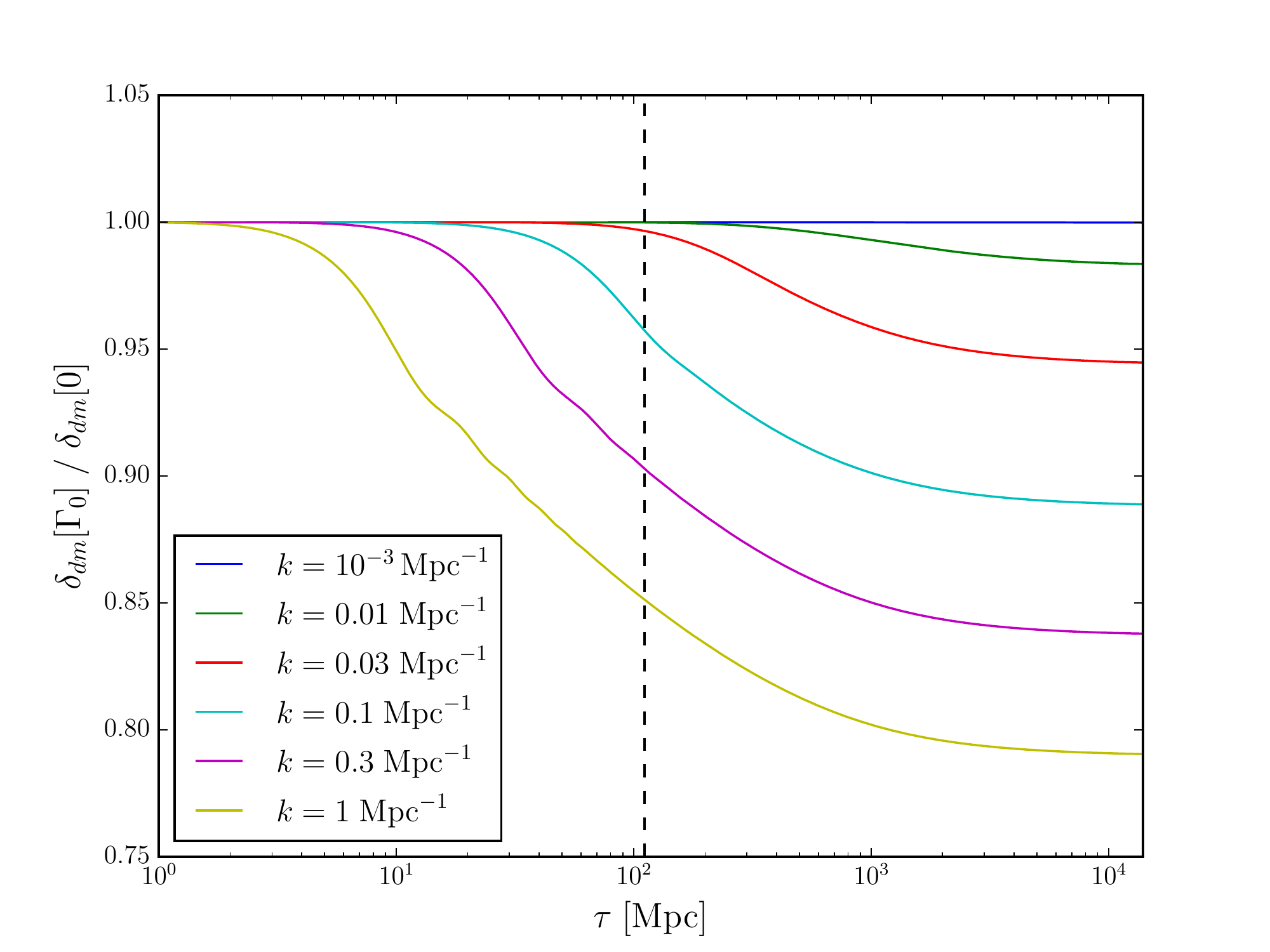}
\caption{\label{fig:growth} Ratio of the dark matter density perturbation $\delta_\mathrm{dm}$ for an interaction rate $\Gamma_0=2 \times 10^{-7}~\mathrm{Mpc}^{-1} \simeq 2 \times 10^{-21}~\mathrm{s}^{-1}$ over the same perturbation in the standard non-interacting limit, in the Newtonian gauge, as a function of conformal time, and for six representative wavenumbers. The interaction rate causes a suppression of  $\delta_\mathrm{dm}$ inside the Hubble radius, efficient especially during radiation domination, and continuing during the beginning of matter domination (the vertical dashed line shows the time of equality between radiation and matter). Apart from $\Gamma_0$, the two cosmological models share the same parameters, including $\Delta N_\mathrm{fluid}=0.21$.
}
\end{figure}

In Section~\ref{sec:fit}, we will find that models with a rate of the order of $\Gamma_0 \simeq 2 \times 10^{-7}~\mathrm{Mpc}^{-1} \simeq 2 \times 10^{-21}~\mathrm{s}^{-1}$ provide the best fits to the data. Figure~\ref{fig:growth} shows the evolution of $\delta_\mathrm{dm}$ for such a value of $\Gamma_0$ normalized to a $\Lambda$CDM model with $\Gamma_0=0$. In the figure, $\Delta N_\mathrm{fluid}=0.21$, but we will later show that the effect of $\Gamma_0$ and $\Delta N_\mathrm{fluid}$ are not strongly correlated. The figure shows the evolution of six different wavenumbers between $k=10^{-3}\, $Mpc$^{-1}$ to $k=1\, $Mpc$^{-1}$. We see that the growth of DM fluctuations is suppressed roughly between $\tau \sim 2\pi/k$ (time of Hubble crossing) and the beginning of matter domination (roughly until $\tau \sim 2500$~Mpc, the time at which the ratio of radiation to matter density is of order 0.1). Later on, i.e. deep in the matter dominated regime and during $\Lambda$ domination, the growth curves are horizontal, showing that the growth rate is the same as in the $\Lambda$CDM model. 

\begin{figure}[thb]
\centering
\includegraphics[width=.49\textwidth]{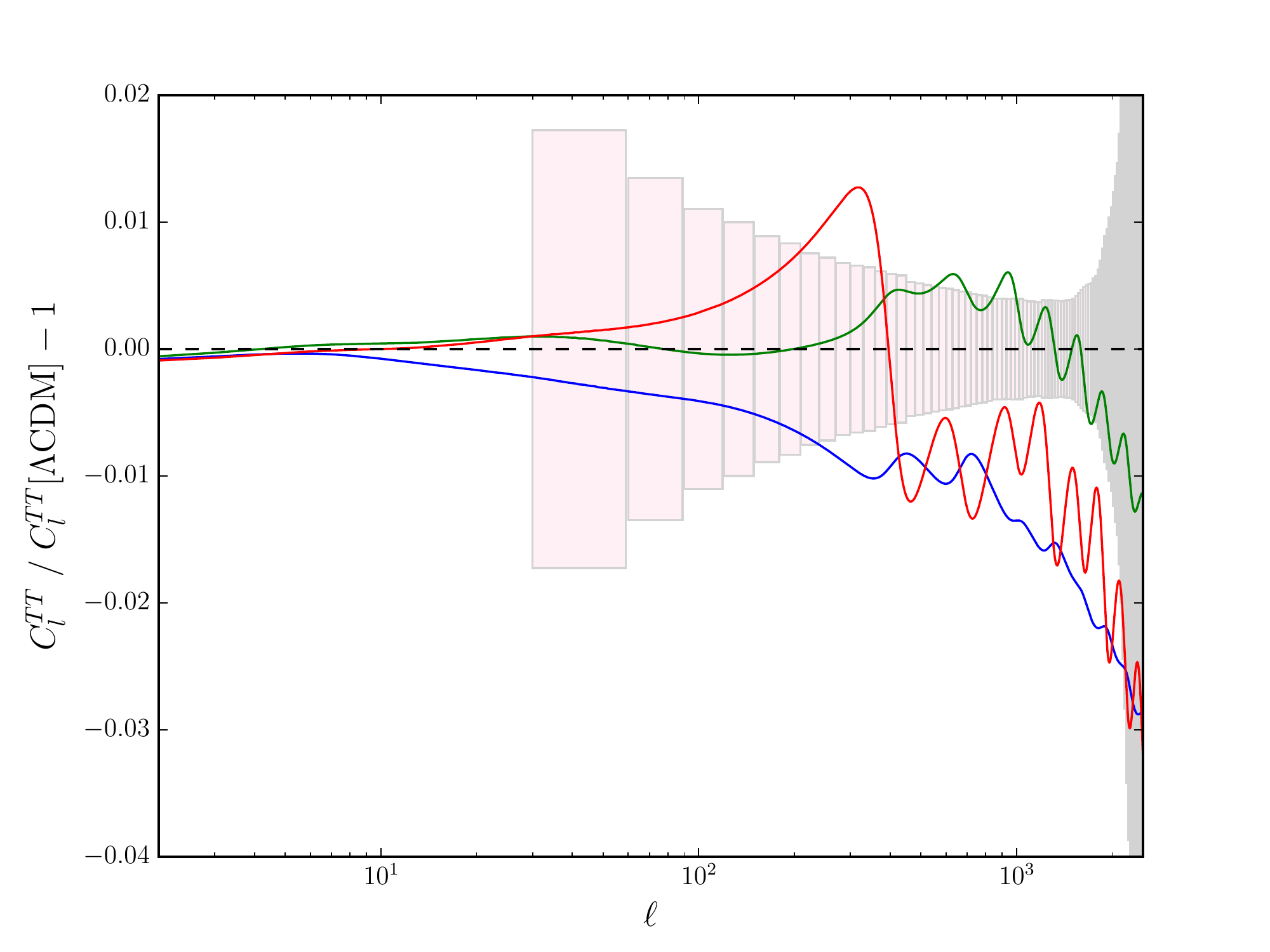}
\includegraphics[width=.49\textwidth]{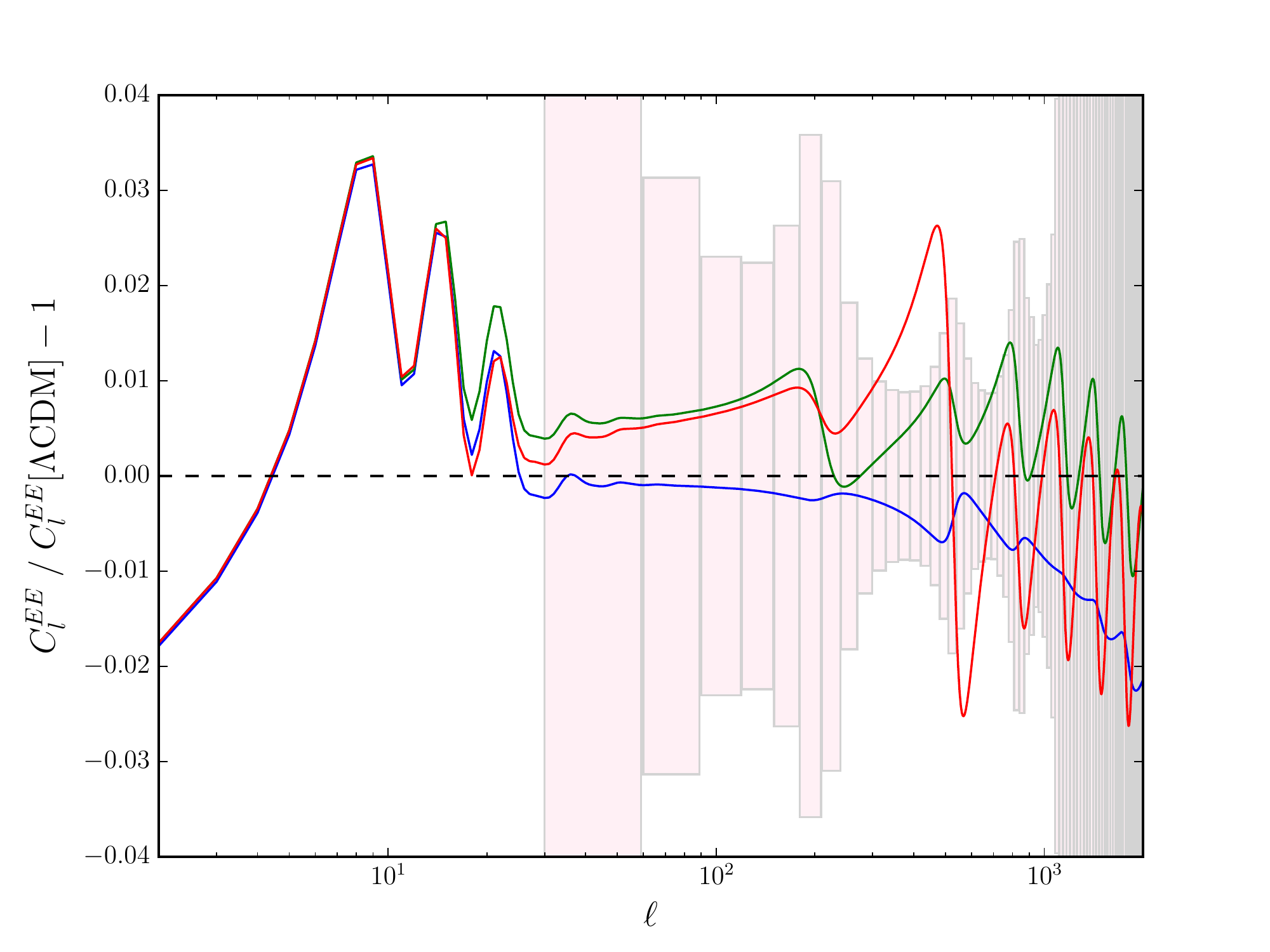}
\caption{\label{fig:CMB} Residual of the temperature (left) and $E$-polarisation (right) power spectrum in several extended models compared to the minimal $\Lambda$CDM model. Two models have ordinary decoupled cold dark matter, but either free-streaming (blue) or self-interacting (green) extra relics with respectively  $\Delta N_\mathrm{eff}=0.21$ or $\Delta N_\mathrm{fluid}=0.21$. The text explains which quantities have been kept fixed in these comparisons. The last model (red curves) shares the same parameters as the latter model (green curves), excepted that the DM-DR interaction is switched on, with $\Gamma_0 = 2 \times 10^{-7}~\mathrm{Mpc}^{-1} \simeq 2 \times 10^{-21}~\mathrm{s}^{-1}$. The boxes show the binned error bars of the Planck High Frequency Instrument 2015 data , which covers $\ell \geq 30$. All models are well within the error bars of the Low Frequency Instrument, which covers $\ell < 30$.}
\end{figure}

\begin{figure}[thb]
\centering
\includegraphics[width=.7\textwidth]{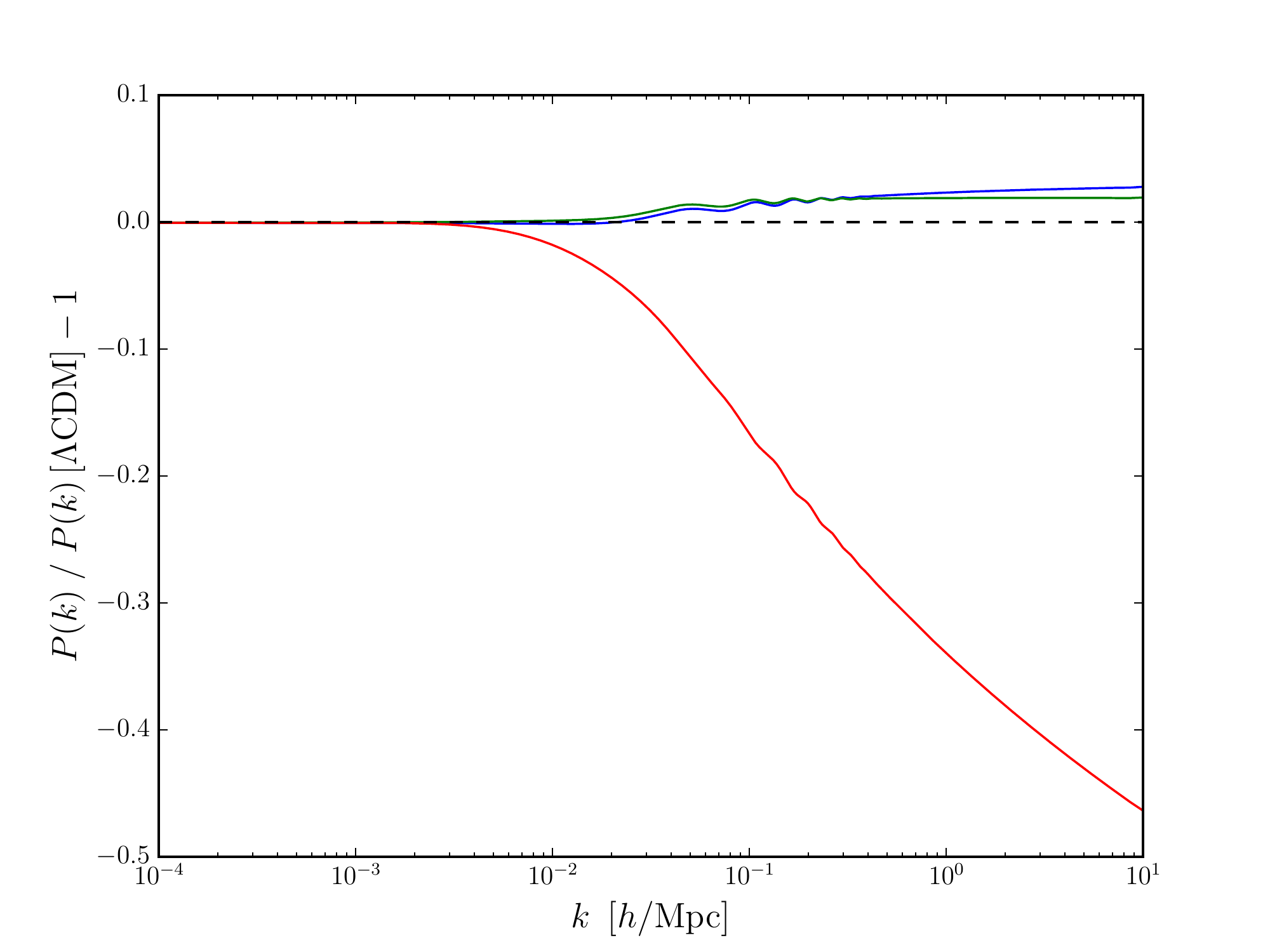}
\caption{\label{fig:PK} Residual of the matter power spectrum $P(k, z=0)$ in the same extended models as in the previous figure, compared to the minimal $\Lambda$CDM model (see the caption of figure~\ref{fig:CMB} for details).}
\end{figure}

Figure~\ref{fig:CMB} shows, first, the effect on the temperature and polarisation CMB spectra of increasing $\Delta N_\mathrm{eff}$ from zero to 0.21, in a model with extra free-streaming relics (e.g. relativistic sterile neutrinos), and in our model with self-interacting dark radiation $\Delta N_\mathrm{fluid}=0.21$.
For a useful comparison, the redshift of radiation/matter and matter/$\Lambda$ equality are kept fixed by appropriately scaling $\omega_\mathrm{dm}$ and $H_0$. The baryon density $\omega_b$ and reionization optical depth are constant. This transformation absorbs a significant part of the total effect, except for perturbation effects (in particular, the gravitational drag exerted on the photons by the DR), and an enhanced Silk damping effect. While the Silk damping effect is the same in the free-streaming and self-interacting models, the perturbation effects are not. In the free-streaming case, the extra radiation has very smooth perturbations, inducing extra damping of the CMB spectra. In the self-interacting case, DR features larger perturbations, which boost the CMB spectra through gravitational coupling. In the latter case, the Silk damping and DR gravitational drag effects tend to compensate each other, and the net variation of the CMB spectrum is smaller than in the free-streaming case. For that reason, we expect a weaker bound on $\Delta N_\mathrm{fluid}$ in the case of self-interacting DR. In both models, the $E$-mode polarisation spectrum is also affected on large angular scales ($l \leq 40$). Indeed, an enhancement of the radiation density changes by a small amount the thermal history (different freeze-out value of the free electron fraction at the end of recombination, and primordial Helium abundance inferred from BBN). In order to maintain the same reionization optical depth, the redshift of reionization changes slightly, and the shape of the low-$l$ polarisation spectrum is affected at the level of a few percents. However, because of cosmic variance, this effect is of small relevance when fitting CMB data.

In Figure~\ref{fig:CMB}, the last (red) curves show the additional effect of switching on the DM-DR interaction, with $\Gamma_0 = 2 \times 10^{-7}~\mathrm{Mpc}^{-1} \simeq 2 \times 10^{-21}~\mathrm{s}^{-1}$, all other parameters being fixed like in the model with self-coupled dark radiation and $\Delta N_\mathrm{fluid}=0.21$ (green curves). The impact of the DM-DR interaction is hence given by the comparison of the red and green curves. Overall, this impact is small, since for the same value of $\Gamma_0$, the CMB is affected at the level of $\sim 2\%$, while the matter power spectrum is affected by 20 to 30\% on the range of scales most relevant for $\sigma_8$. This follows from the fact that in any model in which DM fluctuations evolve on a time scale set by the Hubble rate (rather than some shorter time scale imposed by microphysics), there is an effective gravitational decoupling between DM and photon fluctuations~\cite{Weinberg:2002kg,Voruz:2013vqa}. Models with DM-DR interactions can sometimes violate this condition, and generate ``dark oscillations''~\cite{Cyr-Racine:2013fsa} with a period $T\ll H^{-1}$. However, in our model and for the range of parameters in which we are interested, dark oscillations remain negligible, as clearly shown by Figure~\ref{fig:growth}. Hence the effective gravitational decoupling still holds in good approximation, and primary CMB fluctuations are weakly affected by modifications in the DM growth rate before photon decoupling. Still, they can be affected by modifications in the DR growth rate. 
When the interaction rate $\Gamma_0$ is switched on, DR clusters differently. This propagates to photons through the usual DR gravitational drag effect, i.e. the gravitational interaction between photons and dark radiation. In the CMB spectra, this type of gravitational dragging effect is effective on scales slightly smaller than the Hubble scale at any given time, and shifts the acoustic peaks in phase and amplitude. This explains most of the oscillatory features visible in the red curves of Figure~\ref{fig:CMB}, that start near $l\sim 200$, i.e. on scales which are slightly smaller than the Hubble rate at the time of photon decoupling. The red curves also encodes smaller effects depending slightly on $\Gamma_0$, like the early ISW effect, and lensing by small scale structures.

Figure~\ref{fig:PK} shows the effect of $\Delta N_\mathrm{fluid}$ and $\Gamma_0$ on the matter power spectrum evaluated today, $P(k,z=0)$, for exactly the same models as  in Figure~\ref{fig:CMB}. As discussed in BMS~\cite{Buen-Abad:2015ova}, scales crossing the Hubble radius during matter domination are unaffected by the drag effect, while smaller scales are reduced. The suppression depends on the time spent by each mode inside the Hubble radius during radiation domination. In this example, in the range of scales contributing to $\sigma_8$ ($k \sim 0.2h\,$Mpc$^{-1}$), the matter spectrum is suppressed by about 20\%, as needed for solving the $\sigma_8$ problem.

\section{Example models and calculation of $N_\mathrm{fluid}$ and $\Gamma_0$\label{sec:models}}

In this Section, we briefly review the model of non-Abelian dark matter and dark radiation, more details can be found in BMS \cite{Buen-Abad:2015ova}. We also define an alternate model with a dark photon and give expressions for $\Delta N_\mathrm{fluid}$ and $\Gamma_0$ for both. 

In the non-Abelian DM-DR model the dark matter is a Dirac fermion which transforms as the neutral component of an $SU(2)$-weak triplet. It's couplings to the Standard Model are identical to the pure ``wino" DM in some supersymmetric models, however our model does not have supersymmetry. The DM particle also transforms in the fundamental ``$N$" representation of an $SU(N)$ dark gauge group. Therefore it interacts with ``dark gluons" of the $SU(N)$ gauge group. For phenomenological reasons we will be interested in dark gauge couplings of order $\alpha_d \simlt 10^{-8}$ and gauge groups of size $N=2,3,4$. Since the dark gauge coupling is so small, the dark gauge interactions do not confine until length scales much larger than the size of the visible universe. Therefore the dark gluons correspond to weakly-interacting massless particles, they form the DR of the model.

In \cite{Buen-Abad:2015ova} it was shown that the DM in this model obtains the correct abundance from thermal freeze-out for masses 1.2 TeV (N=2), 1.0 TeV (N=3) and 0.9 TeV (N=4). The model safely evades current DM direct detection bounds, it will be within reach of future indirect detection experiments, and its DM could be discovered at a future 100 TeV collider. Since the dark gluons are so weakly coupled they do not play a direct role in DM detection phenomenology. However, at temperatures above the DM mass, the dark gluons come to thermal equilibrium with the Standard Model so that they obtain the same temperature. After DM annihilates into Standard Model particles and freezes out the dark gluons decouple from the SM and evolve with their own temperature. The temperature of the DR, $T_d$, may be estimated from entropy conservation as $T_d/T_\gamma =(g_*/g_*^{dec})^{1/3}$ where $g_*$ is the effective number of Standard Model degrees of freedom and $g_*^{dec}$ is the number of SM degrees of freedom during the decoupling of the dark radiation.

Using this relationship and assuming that the DR decoupled with temperatures in the 10-80 GeV range, we compare the energy density in the DR fluid with the energy density in standard neutrinos to obtain the effective number of neutrinos \cite{Buen-Abad:2015ova}
\bea
\Delta N_\mathrm{fluid} = 0.07 (N^2-1) \ .
\eea
An important difference between radiation in the form of neutrinos and the dark gluons is that the dark gluons can be described as a perfect fluid with zero viscosity. This is because the rate of self-interactions of gluons - even for couplings as small as $\alpha_d \sim 10^{-10}$ - is faster than the Hubble rate at any time from before nucleosynthesis until today. 

The drag coefficient $\Gamma$ is computed by considering scattering of dark gluons off the dark matter. The leading logarithmically enhanced contribution comes from a t-channel Feynman diagram and was computed in \cite{Buen-Abad:2015ova}. By resumming hard thermal loops the calculation can be improved to also obtain finite pieces, the results are in \cite{Braaten:1991jj, Braaten:1991we}. Here, we only need the logarithmically enhanced contribution 
\bea
\Gamma_0&=&\left. (N^2\!-\!1) \frac{\pi}{9} \alpha_d^2 \log\alpha_d^{-1} \, \frac{T_d^2}{M_\chi} \right|_{today}  \\
&\simeq & 1.9\times 10^{-7}\, {\rm Mpc^{-1}}\, \left[\frac{N^2\!-\!1}{3}\right]\left[ \frac{- \alpha_d^2 \log\alpha_d}{2.0 \times 10^{-16}} \right]\left[\frac{\rm 1.2\, {\rm TeV}}{M_\chi} \right]
\eea
where in the last line we plugged in representative values for the parameters, $N=2$, $\alpha_d=10^{-8.5}$, $M_\chi=1.2$ TeV, and converted to units of inverse megaparsec.

We close this Section by giving an alternate model for DR with couplings to the DM which predicts values for $\Delta N_\mathrm{fluid}$ as low as 0.07. In this model the DM is a Dirac fermion wich couples to a massless dark photon, the gauge boson of a dark $U(1)$ gauge group. In addition to the DM and the dark photon the model also includes massless fermions which are charged under the dark $U(1)$ (in an anomaly free representation). For simplicity, we consider a single Dirac DM particle with charge $1$ and $N_f$ species of massless Dirac fermions with charge $q_\mathrm{dr}$. The DM is assumed to couple to the Standard Model through the weak interactions or through the Higgs portal. Then it can obtain its abundance from standard thermal freeze-out (\ie it is a WIMP) with a mass of order the weak scale. The massless fermions and the dark photon constitute the DR of the model. The dark photon equilibrates with the Standard Model at temperatures above the DM mass for the range of gauge couplings of interest. However, whether or not the light fermions equilibrate as well depends on the precise values of their charges. We find that for $q_\mathrm{dr} \simlt 1/3$ the coupling of the light fermions to the SM thermal bath are too weak to bring them to thermal equilibrium before the decoupling of the dark sector from the SM. In this case they could remain significantly colder than the SM until after the decoupling of the dark photons from the SM. Then they would give a negligible  contribution to the effective number of radiation degrees of freedom. However, for light fermion charges greater than 1 the light fermions equilibrate before decoupling of the dark radiation. After DM freeze-out the interactions between the SM and the DR become too infrequent to maintain thermal equilibrium and the two fluids decouple.  As in the non-Abelian model, the DR is described by a perfect fluid with zero viscosity at all temperatures relevant for the growth of density perturbations and the CMB even in the case of smaller light fermion charges. 

We obtain 
\bea
\Delta N_\mathrm{fluid} = 0.07 \,(1+\frac{7}{4}N_f) 
\eea
for the case of $q_\mathrm{dr} \simgt 1$ and 
\bea
\Delta N_\mathrm{fluid} = 0.07  
\eea
for $q_\mathrm{dr} \simlt 1/3$, and values of $N_\mathrm{fluid}$ ranging between the two limiting cases for  charges $1 \simlt q_\mathrm{dr} \simlt 1/3$. For the drag coefficient we obtain
\bea
\Gamma_0 &=&\left. N_f\,  q_\mathrm{dr}^2 \frac{2\pi}{9} \alpha_d^2 \log\alpha_d^{-1} \, \frac{T_d^2}{M_\chi} \right|_{today}  \\ 
&\simeq & 1.8\times 10^{-7}\, {\rm Mpc^{-1}}\,\left[\frac{N_f\,  q_\mathrm{dr}^2}{2}\right]\left[ \frac{- \alpha_d^2 \log\alpha_d}{2.0 \times 10^{-16}} \right]\left[\frac{\rm 1.7\, {\rm TeV}}{M_\chi} \right].
\eea

\section{Fit to current data\label{sec:fit}}

\subsection{Data and methodology\label{subsec:data}}

We use the code {\sc MontePython}~\cite{Audren:2012wb} to fit the model to currently available cosmological data. We split the data into four categories:
\begin{itemize}
\item {\bf CMB}: we use the Planck 2015 TT + low-$\ell$ likelihood from Ref.~\cite{Aghanim:2015xee}.
\item {\bf BAO}: we use measurements of $D_V/r_\mathrm{drag}$ at $z=0.106$ by 6dFGS~\cite{2011MNRAS.416.3017B}, at $z=0.15$ by SDSS-MGS~\cite{Ross:2014qpa}, at $z=0.32$ by BOSS- LOWZ~\cite{Anderson:2013zyy}, and anisotropic BAO measurements at $z=0.57$ by BOSS-CMASS-DR11~\cite{Anderson:2013zyy}.  
\item {\bf LSS}: we use three probes of Large Scale Structure: the Planck 2015 lensing likelihood~\cite{Ade:2015zua}, the constraint $\sigma_8 (\Omega_m/0.27)^{0.46} = 0.774 \pm 0.040$ (68\%CL) derived from the weak lensing survey CFHTLenS~\cite{Heymans:2013fya}, and the constraint $\sigma_8 (\Omega_m/0.27)^{0.30} = 0.782 \pm 0.010$ (68\%CL) from Planck SZ cluster mass function~\cite{Ade:2013lmv}. The latter constraints should be taken with a grain of salt, because they have been inferred under the assumption of a $\Lambda$CDM model. However, our model produces a featureless matter power spectrum on the scales probed by these experiments, so these constraints are probably valid to a good approximation.
\item {\bf H$_0$}: we occasionally also use the constraint $H_0=73.8\pm2.4$~km/s/Mpc (68\%CL) from Riess et al. \cite{Riess:2011yx}. Direct measurements of the local Hubble rate by e.g. \cite{Riess:2011yx,Freedman:2012ny} have been questioned recently by the community, with the concern that systematic errors might have been underestimated. However we will use it only in order to show that our model is well compatible with such high values of the Hubble rate.
\end{itemize}
We did several MCMC runs with various combinations of these data sets, for the 6-parameter $\Lambda$CDM model, and for our 8-parameter model (with a free effective number of dark gluons $\Delta N_\mathrm{fluid}$, and a dark matter-dark gluon interaction rate $\Gamma_0$, expressed in the code in inverse Mega-parsecs). 
We define the $\Lambda$CDM in the same way as the ``base model'' in Planck 2013 and Planck 2015, including two massless and one massive neutrino species with $m=0.06$~eV, and assuming an effective neutrino number $N_\mathrm{eff}=3.046$. We keep exactly the same settings in the model with interacting dark matter and dark gluons; in that case, the density of usual CDM is set to zero, while the density of non-abelian dark matter is parametrized by $\omega_\mathrm{dm}=\Omega_\mathrm{dm}h^2$.

We impose flat priors on the parameters of our model: $\{ \omega_b$, $\omega_\mathrm{dm}$, $\Delta N_\mathrm{fluid}$, $\Gamma_0$, $H_0$, $A_s$,  $n_s$, $\tau_\mathrm{reio}\}$. Only the lower edge of the priors on $\Delta N_\mathrm{fluid}$, $\Gamma_0$ and $\tau_\mathrm{reio}$ are relevant. For the interaction rate, we just require $\Gamma_0\geq0$. We impose a  prior $\tau_\mathrm{reio}\geq0.04$ on the optical depth to reionisation\footnote{This is done in order to avoid the limit $\tau_\mathrm{reio}\longrightarrow 0$, which is unphysical given the residual ionisation fraction after recombination.}. Finally, for the dark gluon density parameter $\Delta N_\mathrm{fluid}$, our non-Abelian DM-DR model predicts discrete values $\Delta N_\mathrm{fluid} = 0.21, 0.56, 1.05, ...$ for $N=2,3,4,...$, while the second model discussed in section~\ref{sec:models} predicts $\Delta N_\mathrm{fluid} = 0.19, 0.32, 0.44, ...$ for $N_f= 1, 2, 3, ...$ in the case of large light fermion charges and $\Delta N_\mathrm{fluid} = 0.07$ for small charges. Hence we stick to the theoretical prior $\Delta N_\mathrm{fluid} \geq 0.07$ in all our MCMC runs. This means that the standard $\Lambda$CDM model (with $\Gamma_0=0$ and $\Delta N_\mathrm{fluid}=0$) is not a special point in the parameter space of our ``dark matter drag'' models. This is not a problem from the point of view of the statistical analysis, as long as we provide a way to evaluate the goodness-of-fit of the ``dark matter drag'' model compared to $\Lambda$CDM. For that purpose, we performed some companion runs with $\Lambda$CDM and the same combinations of data; in each case, we report the difference between the minimum $\chi^2$ of the two models\footnote{In principle, if we were using the Multinest algorithm instead of the Metropolis-Hastings algorithm, we could also report the Bayesian evidence ratio between the two models. This is beyond the scope of this paper.}.

Exploring the range [0-0.07] could be theoretically motivated by assuming some entropy production mechanism, like the decay of some other particles into the SM thermal bath. In that case, more ingredients are needed, and physical effects on the CMB and LSS observables are a bit different. Indeed, in the limit of very small DR density, one can reach a new regime in which the effect of the drag of DM on DR can be as relevant as the drag of DR on DM. We defer the study of this other class of models to a future publication.

\begin{table}[th]
\begin{tabular}{|c|c|c|c|c|}
\hline
Parameter & CMB+BAO & CMB+LSS & CMB+BAO & CMB+BAO \\
                  &                    &                  &     +LSS      & +LSS+$H_0$ \\
\hline
$100\omega_b$                          & $2.236_{-0.026}^{+0.024}$ & $2.219_{-0.041}^{+0.029}$ & $2.220_{-0.025}^{+0.021}$ &  $2.234_{-0.026}^{+0.025}$ \\
$\omega_\mathrm{dm}$             & $0.1244_{-0.0040}^{+0.0021}$ & $0.1256_{-0.0047}^{+0.0034}$ & $0.1249_{-0.0049}^{+0.0023}$ &  $0.1274_{-0.0060}^{+0.0040}$ \\
$\Delta N_\mathrm{fluid}$             & $<0.58$ & $<0.71$  & $<0.67$ &  $<0.59$ \\
$10^7 \Gamma_0$ [Mpc$^{-1}$]   & $<1.54$ & $1.74_{-0.55}^{+0.57}$ &  $1.65_{-0.44}^{+0.42}$ & $1.69_{-0.48}^{+0.43}$ \\
$H_0$  [km/(s Mpc)]                                  & $69.1_{-1.3}^{+0.8}$ & $69.0_{-2.4}^{+1.4}$ & $69.1_{-1.5}^{+0.8}$ & $70.2_{-1.6}^{+1.3}$ \\
$10^9 A_s$                          & $2.220_{-0.081}^{+0.079}$ & $2.205_{-0.076}^{+0.063}$ & $2.205_{-0.069}^{+0.063}$ & $2.217_{-0.070}^{+0.062}$ \\
$n_s$                                   & $0.9709_{-0.0053}^{+0.0048}$ & $0.9762_{-0.0081}^{+0.0070}$ & $0.9736_{-0.0055}^{+0.0051}$ &  $0.9796_{-0.0053}^{+0.0049}$ \\
$\tau_\mathrm{reio}$           & $0.084_{-0.019}^{+0.018}$ & $0.078_{-0.019}^{+0.016}$ & $0.079_{-0.015}^{+0.015}$ & $0.082_{-0.016}^{+0.014}$ \\
\hline   
$\Omega_\mathrm{m}$       & $0.3088_{-0.0083}^{+0.0082}$ & $0.3130_{-0.018}^{+0.019}$ & $0.3097_{-0.0083}^{+0.0085}$ & $0.3052_{-0.0083}^{+0.0080}$ \\
$\sigma_8$                          & $0.811_{-0.019}^{+0.026}$ & $0.760_{-0.019}^{+0.017}$ & $0.762_{-0.011}^{+0.011}$ & $0.766_{-0.011}^{+0.011}$ \\
\hline
$\Delta \chi^2$ / $\Lambda$CDM & 0 & -9.6 & -11.4 & -12.7 \\
\hline
\end{tabular}
\caption{\label{tab1}
Mean value and 68\%CL confidence interval (or, in a few cases, 95\%CL upper limit) for the eight parameters of our model (assuming flat priors) and two derived parameters. The last line shows the minimum $\chi^2$ value compared to that of $\Lambda$CDM with the same data. Note that $10^7 \Gamma_0$ in Mpc$^{-1}$ is equal to $10^{21} \Gamma_0$ in s$^{-1}$ to 3\% accuracy.
}
\end{table}

\subsection{Reconciling cosmological data sets}

Our results are summarized by Table~\ref{tab1}. The most striking facts are, first, that our model can reconcile CMB, BAO and LSS data, and even the $H_0$ measurement of \cite{Riess:2011yx}; and second, that when at least CMB and LSS data are included in the fit, the minimum effective $\chi^2$ decreases by a substantial amount when going from the $\Lambda$CDM model to our model: $\Delta \chi^2=-9.6$ for CMB+LSS and $\Delta \chi^2=-11.4$ for CMB+BAO+LSS.

\begin{figure}[thb]
\centering
\includegraphics[width=.32\textwidth]{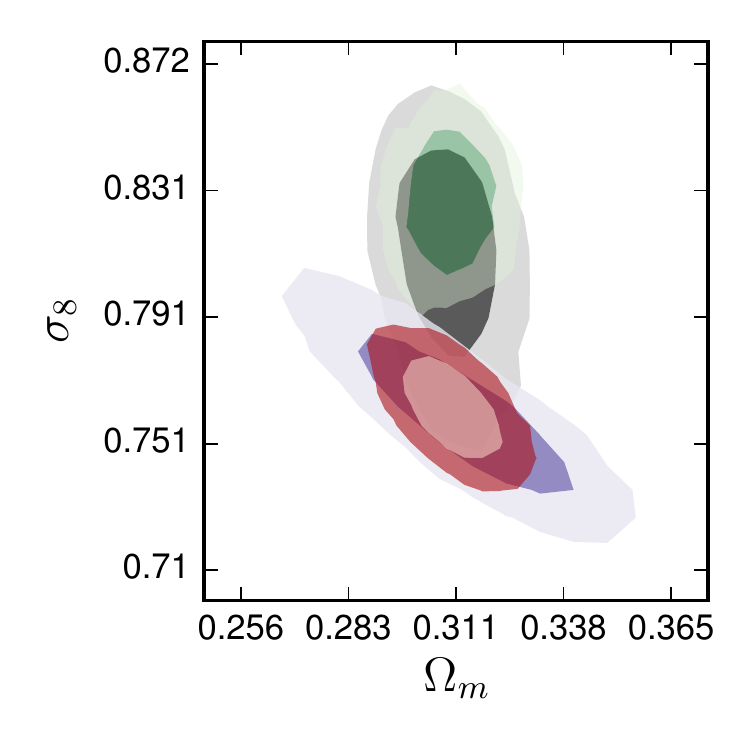}
\includegraphics[width=.32\textwidth]{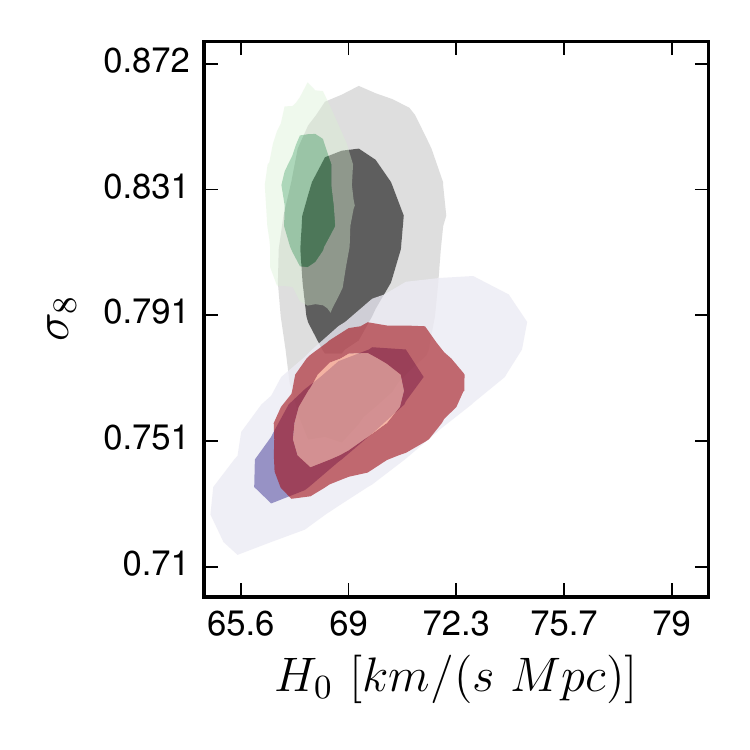}
\caption{\label{fig:sigma8} 68\% and 95\% CL contours for ($\sigma_8, H_0$) and ($\sigma_8, \Omega_m$): first, for the $\Lambda$CDM model and CMB+BAO data (green); next, for our model and CMB+BAO data (black), CMB+LSS data (blue), CMB+BAO+LSS data (red). This figure can be compared with Fig.~33 of Planck 2015 \cite{Planck:2015xua}, to show a clear difference between our model and all the massive active/sterile neutrino models used in that figure: our model can explain a lower $\sigma_8$ without requiring at the same time a lower $H_0$ or a higher $\Omega_m$ (on the contrary, it is compatible with higher $H_0$ values).}
\end{figure}

A good way to appreciate these results is to look at the ($\sigma_8, H_0$) and ($\sigma_8, \Omega_m$) contours shown in Figure~\ref{fig:sigma8}. The CMB+BAO results for $\Lambda$CDM are shown in green. These results are notoriously in 3-4$\sigma$ tension with LSS data, which require at the same time a lower $\sigma_8$ and a similar $\Omega_m$, and in 2-3$\sigma$ tension with the high value of $H_0$ from \cite{Riess:2011yx}. The CMB+BAO results for our model are shown in black/grey. The comparison of the green and black contours makes the point. Our model  is compatible with much lower values of $\sigma_8$ {\it for the same range of $\Omega_m$ values}. It is also compatible with much larger $H_0$ values. It is worth stressing that this represents a crucial difference between our model and more traditional models featuring extra relativistic or massive relics (like sterile neutrinos) in combination with massive active neutrinos. These models have been invoked by Ref.~\cite{Hamann:2013iba} to reconcile tensions between CMB, LSS and $H_0$ data. The Planck 2015 paper has shown that this does not work well anymore with recent CMB and BAO data. Ref.~\cite{Planck:2015xua} shows that the improvement is only of the order of $\Delta \chi^2\sim 1$ between the $\Lambda$CDM model and these models.  Fig.~33 of \cite{Planck:2015xua} provides a clear interpretation of this result. In models with extra massless/massive neutrinos, parameter correlations are such that a reduction of $\sigma_8$ requires higher values of $\Omega_m$ and smaller values $H_0$, which exacerbates tensions. We emphasise that this is not the case in our model, as can be seen from the black contours in Figure~\ref{fig:sigma8}: the effect of the interaction rate allows for a lower $\sigma_8$ without correlated effects on $\Omega_m$, and $H_0$. 
Hence we do obtain a substantial improvement in $\chi^2$.

The blue contours in Figure~\ref{fig:sigma8} correspond to CMB+LSS results for our model. A discrepancy between black and blue contours would reveal a tension between BAO and LSS data (given the CMB data). Most extensions of the minimal $\Lambda$CDM model studied so far lead to such a tension.
Instead, for our model, the joint two-dimensional contours overlap at the 1.1$\sigma$ level, and the 1-dimensional posteriors for $\sigma_8$ are compatible at the 1.4$\sigma$ level.
This shows that our model actually reconciles CMB, BAO and LSS data. 

The joint contours for CMB+BAO+LSS are shown in red. 
Note that they are $\sim$3$\sigma$ 
away from the green contours of the $\Lambda$CDM model, which is consistent with the typical level of tension between CMB+BAO and LSS data in the $\Lambda$CDM case, and with the fact that $\Delta \chi^2=-11.4$ for CMB+LSS+BAO. The preferred parameter values for this data set are $\sigma_8=0.762_{-0.011}^{+0.011}$, $\Omega_m=0.3097_{-0.0083}^{+0.0085}$, $H_0=69.1_{-1.5}^{+0.8}$ km/s/Mpc, all at 68\%CL. The latter result on $H_0$ is compatible with the measurement of \cite{Riess:2011yx} at the 1$\sigma$ level. Hence it is legitimate to combine CMB+BAO+LSS data with this measurement. In that case, the minimum $\chi^2$ decreases by 12.7 with respect to $\Lambda$CDM.

\subsection{Physical interpretation of the results}

\begin{figure}[thb]
\centering
\includegraphics[width=.9\textwidth]{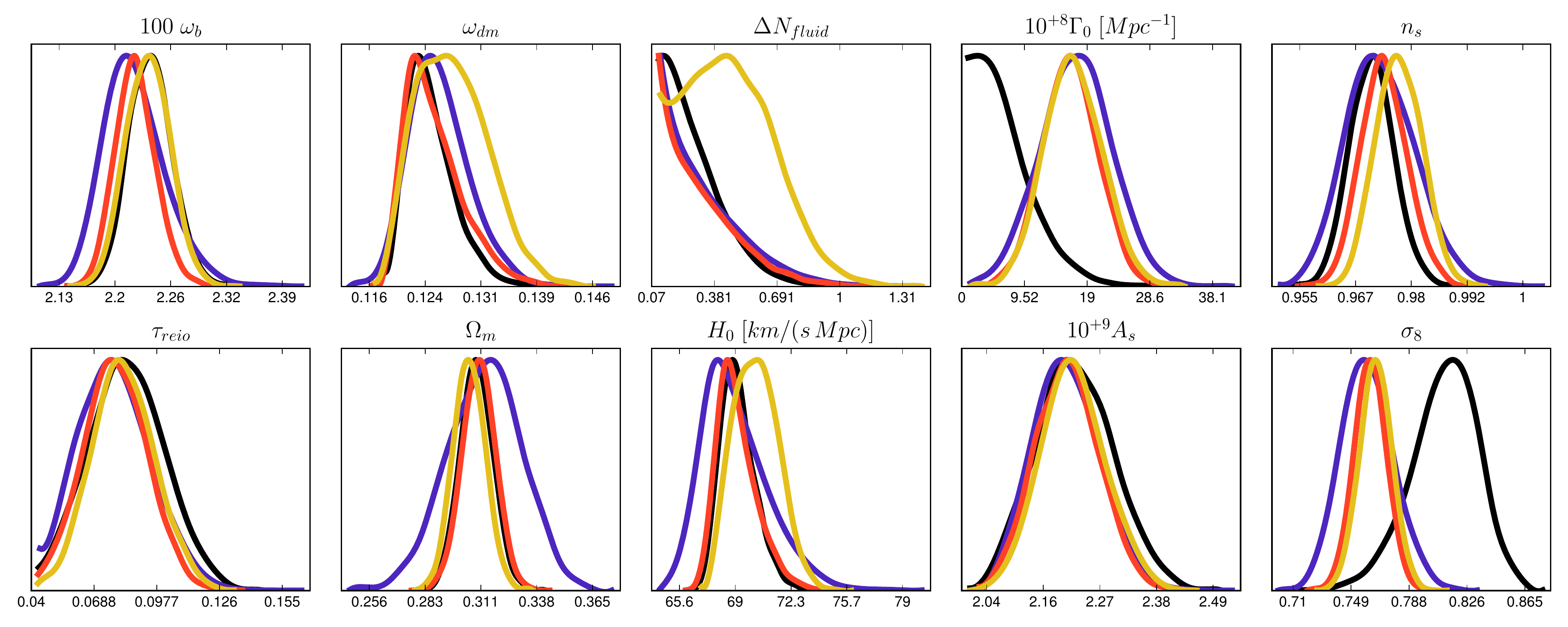}
\caption{\label{fig:1d} Posterior probabilities for the eight parameters forming the basis of our model and for two derived parameters ($\Omega_m$, $\sigma_8$), for CMB data combined with BAOs (black), LSS (blue), BAO+LSS (red), BAO+LSS+$H_0$ (yellow). See the text for details on parameter definitions and priors, and for the precise content of each dataset.}
\end{figure}

Figure~\ref{fig:1d} shows the posterior probability of the parameters of our model for the four combinations of data that we investigated. Since the dark matter--dark radiation interaction rate suppresses the matter power spectrum on small scales, this rate is compatible with zero for CMB+BAO data, and 3-4$\sigma$ away from zero as soon as LSS data is introduced. The figure also shows that adding an $H_0$ prior has very little effect, excepted on $H_0$ itself, 
and also on $\Delta N_\mathrm{fluid}$, due to the correlation between these two parameters discussed in the next paragraph.

\begin{figure}[thb]
\centering
\includegraphics[width=.32\textwidth]{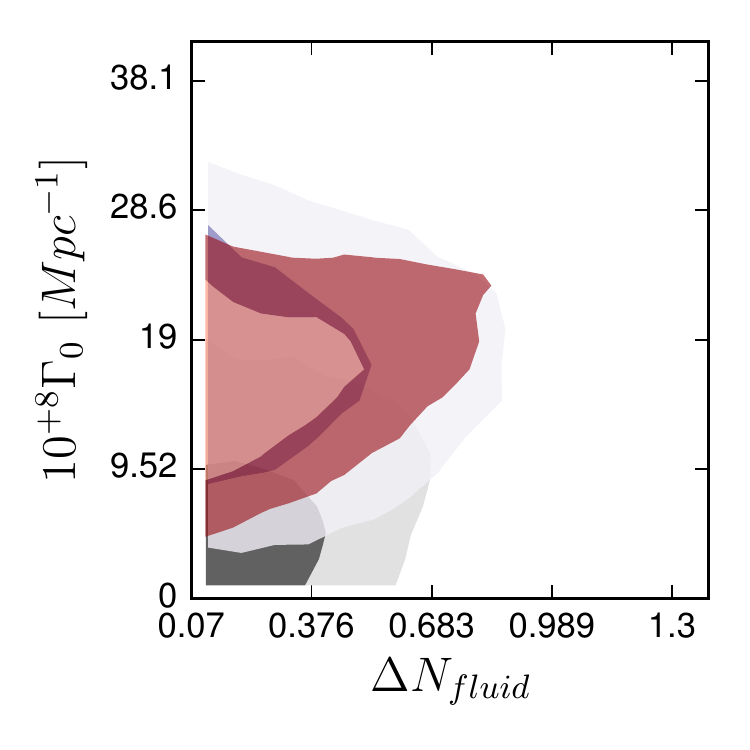}
\includegraphics[width=.32\textwidth]{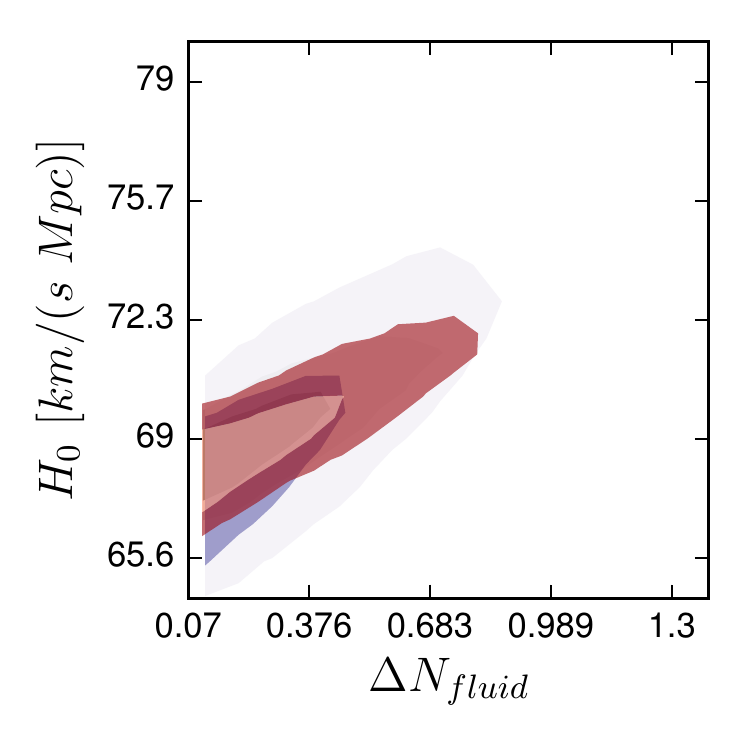}
\includegraphics[width=.32\textwidth]{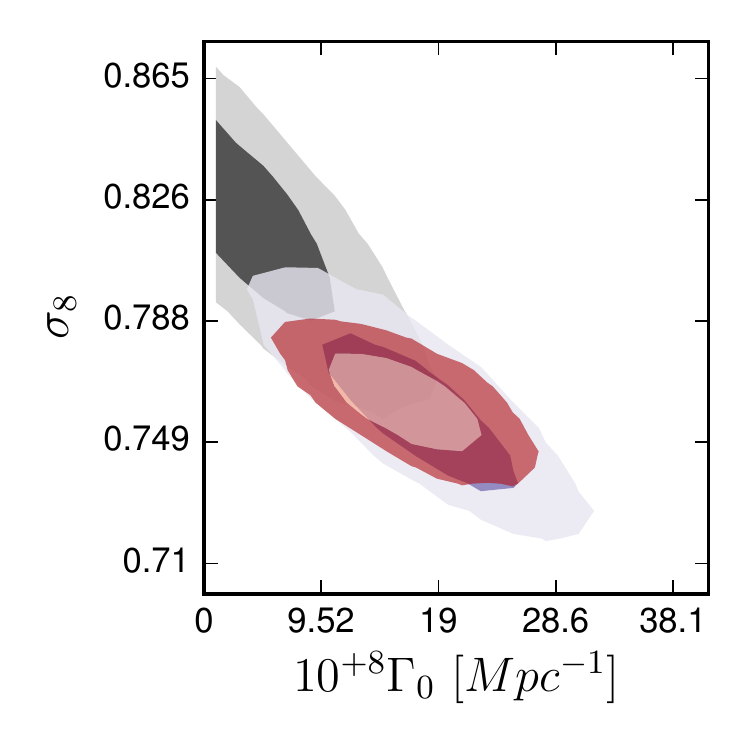}
\caption{\label{fig:2d} 68\% and 95\% CL contours for ($\Delta N_\mathrm{fluid}$, $\Gamma_0$), ($\Delta N_\mathrm{fluid}$, $H_0$), ($\Gamma_0$, $\sigma_8$), with CMB+BAO data (black), CMB+LSS data (blue), CMB+BAO+LSS data (red).}
\end{figure}

The relation between the observable parameters $(\sigma_8, H_0)$ and the fundamental parameters $(\Gamma_0, \Delta N_\mathrm{fluid})$ is better illustrated by figure~\ref{fig:2d}. The 
fact that the dark matter--dark radiation interaction has the effect of reducing the small-scale matter power spectrum is directly responsible for the strong correlation between $\sigma_8$ and $\Gamma_0$. Concerning the correlation between $H_0$ and $\Delta N_\mathrm{fluid}$, a few comments are in order. At the level of background cosmology, $\Delta N_\mathrm{fluid}$ has a model independent effect (similar for our interacting dark radiation or for ordinary decoupled relativistic relics, like very light sterile neutrinos). It is well-known that  an increase in the radiation density is compatible with CMB+BAO data provided that the characteristic redshifts $z_\mathrm{eq}$, $z_\Lambda$ of radiation/matter and matter/$\Lambda$ equality remain constant, and in our basis of parameters, this means that $H_0$ must increase. This leads to a parameter degeneracy between $\Delta N_\mathrm{fluid}$ and $H_0$, that we clearly see in figure~\ref{fig:2d}. This degeneracy is not perfect,  due to an enhancement of the Silk damping effect when $H_0$ increases, and to more subtle perturbation effects. The latter effect is more model dependent: perturbation effects are less pronounced for interacting dark radiation, since it does not free-stream like extra massless relics. 
Nevertheless, we find bounds on $\Delta N_\mathrm{fluid}$ for our dark radiation model which are comparable to those in the case of standard extra relativistic relics. For instance, with CMB+BAO data, we get
$\Delta N_\mathrm{fluid}<0.58$ (95\%CL);  while with the same CMB and BAO data, Planck 2015~\cite{Planck:2015xua} found $\Delta N_\mathrm{eff}<0.61$ (95\%CL) (with a different prior, equivalent to $-3<\Delta N_\mathrm{eff}<3$, instead of our prior $\Delta N_\mathrm{fluid}>0.07$). 

Note that Planck data is more compatible with the presence of three standard free-streaming neutrinos than with an equivalent amount of radiation with effective parameters $(c_s^2, c_\mathrm{vis}^2) \neq (1/3, 1/3)$~\cite{Gerbino:2013ova,Audren:2014lsa,Planck:2015xua}. This just means that we do observe the free-streaming effect of the SM active neutrinos in the CMB spectrum. But when we allow for extra species, we can accommodate some amount of dark radiation even if it does not free stream.

In summary, in our fits of the data, we observe two simple and independent effects: the rate $\Gamma_0$ helps to reduce $\sigma_8$, while the relic density of dark radiation, parametrized by $\Delta N_\mathrm{fluid}$, helps to increase $H_0$. In the class of models studied here, which have $\Delta N_\mathrm{fluid}\geq0.07$, there is no obvious correlation between these two effects, as shown by the left plot in Figure~\ref{fig:2d}. The interaction rate can efficiently suppress $\sigma_8$ even when $\Delta N_\mathrm{fluid}=0.07$, the minimum attainable value in our particle physics models. 
The best fit without $H_0$ prior corresponds to the Abelian model in which the light fermions do not thermalize before the dark sector decouples. The best fit with the $H_0$ prior is close to the $SU(3)$ non-Abelian model, and in any case $SU(N)$ with $N=4$ or higher  is clearly disfavored. Since there is no significant correlation between the effects of $\Gamma_0$ and $\Delta N_\mathrm{fluid}$, we get evidence for a non-zero interaction rate solely from LSS data, independently of the value of $H_0$, which is still subject to strong observational uncertainties.

\section{Conclusions}

In this paper we showed how models of dark matter interacting with dark radiation can reconcile CMB and LSS data. This requires a smooth reduction in the growth of modes that entered the horizon before matter domination, this includes the modes associated with the characteristic scales relevant to $\sigma_8$. We showed that such a smooth reduction of power arises if the momentum transfer between DR and DM is very small and scales as $T^2$, such that it can be equally important throughout the radiation dominated epoch but small enough to avoid dark acoustic oscillations. To the best of our knowledge the class of models studied in this paper are the first to reconcile LSS data with CMB without degrading the goodness of the fit to the latter. 

We have modified the Boltzmann code CLASS to include interacting DM and DR, and parametrized the energy density in DR by the equivalent effective number of neutrinos, $\Delta N_\mathrm{fluid}$, and the interaction strength by the momentum transfer today $\Gamma_0$. Motivated by particle physics models in which dark matter and dark radiation were in thermal equilibrium with the Standard Model in the early universe we have focused on a minimum value of $\Delta N_\mathrm{fluid} > 0.07$.

We use the code {\sc MontePython}~\cite{Audren:2012wb} to fit the model to CMB, BAO and LSS data and found that our model improves $\chi^2$ by 11.4
compared to $\Lambda$CDM. The best fit has $\Gamma_0 = \left( 1.65_{-0.44}^{+0.42} \right) 10^{-7} \, \text{Mpc}^{-1} (\sim 1.6 \times 10^{-21} \, \text{s}^{-1}$), $\Omega_m= 0.3097_{-0.0083}^{+0.0085}$, $H_0=69.1_{-1.5}^{+0.8}$ km/s/Mpc, all at 68\%CL, and $\Delta N_\mathrm{fluid}<0.67$ at 95\%CL. We find $3-4 \, \sigma$ evidence for non-zero momentum transfer rate between DM and DR. We also find that our model can accommodate larger values of $H_0$, which have been reported by direct measurements~\cite{Riess:2011yx}. This is a significant improvement over $\Lambda$CDM, for which an increase in $H_0$ exacerbates the tension with LSS data because it leads to an even larger prediction for $\sigma_8$.

Dark matter drag also predicts a smooth suppression of the matter power spectrum at scales smaller than the $\sim 8$ Mpc relevant for $\sigma_8$. Thus there may be observable effects from the Lyman-$\alpha$ forest and perhaps also at even smaller scales where there are difficulties with reconciling observations and simulations of DM halos (see \cite{Seljak:2006bg,Viel:2013fqw,Palanque-Delabrouille:2015pga} and also \cite{Weinberg:2013aya,Brooks:2014qya}).

In the near future there will be a wealth of new data, especially for LSS (DES, LSST, Euclid, etc.). If the new data sets continue to favor an interacting DM and DR scenario over $\Lambda$CDM, this could be the first clear experimental evidence for non-gravitational interactions of DM.

\section{Acknowledgments}
We wish to thank Joanne Cohn, David Kaplan, Lloyd Knox and Martin White for useful discussions. We acknowledge the use of data produced by the {\sc planck} collaboration. The work of GMT and MS is supported by the US Department of Energy Office of Science under Award Number DE-SC-0010025. GMT also acknowledges support from a DOE High Energy Physics Graduate Fellowship. The computational work for this paper was performed on the Shared Computing Cluster which is administered by Boston University’s Research Computing Services.

%%%%%%%%%%%%%%%%%%%%%%%%%%%%%%%%%%%%%%%%%%%%%%%%%%%%%%%%%%%%

\bibliography{lms-dm-interactions}{}

\end{document}